# Prediction of Memory Retrieval Performance Using Ear-EEG Signals

Jenifer Kalafatovich, Minji Lee, and Seong-Whan Lee, *Fellow*, *IEEE*

*Abstract*— Many studies have explored brain signals during the performance of a memory task to predict later remembered items. However, prediction methods are still poorly used in real life and are not practical due to the use of electroencephalography (EEG) recorded from the scalp. Ear-EEG has been recently used to measure brain signals due to its flexibility when applying it to real world environments. In this study, we attempt to predict whether a shown stimulus is going to be remembered or forgotten using ear-EEG and compared its performance with scalp-EEG. Our results showed that there was no significant difference between ear-EEG and scalp-EEG. In addition, the higher prediction accuracy was obtained using a convolutional neural network (pre-stimulus: 74.06%, on-going stimulus: 69.53%) and it was compared to other baseline methods. These results showed that it is possible to predict performance of a memory task using ear-EEG signals and it could be used for predicting memory retrieval in a practical brain-computer interface.

*Clinical Relevance*— We proposed a more flexible and practical method to predict memory retrieval performance using brain signals, specifically those around the ear.

## I. INTRODUCTION

Memory is a vital process in daily life, it allows to store new information, and retrieve it later [1]. Neural changes during the presentation of a stimulus that characterized the later remembering or forgetting are known as subsequent memory effects [2]. These changes have been studied before using functional magnetic resonance imaging (fMRI), intracranial electroencephalogram (ECoG), and electro-encephalogram (EEG) [3, 4] for understanding memory processes and recently for the prediction of subsequent memory effects [5-7]; however, achieved prediction accuracy still low, especially when using EEG signals.

EEG has been widely explored to measure brain activity from scalp and applied to different brain-computer interfaces (BCI) paradigms such as steady-state visual evoked potentials [8], evoked related potentials (ERPs) [9]; and conscious monitoring [10]. Its advantages over other methods to measure brain activity (fMRI, ECoG, etc) are its high temporal resolution and cheapness. In studies related to memory, EEG analysis had shown significant changes in frequency band and ERP. Hanslmayr et al. [11] reported an increase in theta frequency band during pre-stimulus related to successful subsequent memory effect. Other studies have found an increase of theta and low beta activity during on-going stimulus over frontal and temporal areas [12, 13]. An ERP study reported negative activity around 250 ms prior to stimulus [14].

Ear-EEG is a noninvasive method to measure brain activity by placing electrodes around the ear instead of scalp [15]. Its use has been increasing due to its advantages over scalp-EEG such as flexibility and portability, which provide a more real and practical method for measuring brain signals. Ear-EEG has been applied to BCI [16], sleep stage classification [17], and fatigue detection [18]. However, reported classification accuracies are still low; mainly due to the long distance of electrodes and task-related brain regions. On the other hand, ear-EEG applied to auditory paradigms has shown great performance due to its proximity to temporal cortex [14].

In this study, we explore the possibility to predict whether a shown stimulus is going to be later remembered or forgotten using ear-EEG. Brain activity during the performance of the encoding phase was recorded and later analyzed for the prediction. Previous studies have used scalp-EEG for studying memory process and reported the importance of the temporal cortex in memory task; therefore, we propose a system for prediction of memory performance using ear-EEG signals. We applied common spectral pattern (CSP) and filter bank CSP (FBCSP) as feature extraction due to the relevance of neural oscillations in memory task, and linear discriminant analysis (LDA) as classification method. Additionally, a convolutional neural network (CNN) was implemented in order to improve the performance of other methods. Finally, we compared the obtained prediction accuracy to find significant differences. We hypothesized that it is possible to predict retrieval performance using ear-EEG with similar accuracy than when using scalp-EEG due to the importance of temporal sites in memory task. Our findings lead to a better prediction when using ear-EEG signals compared to other methods; therefore, it opens the possibility for the implementation of a more flexible way to predict and enhance memory abilities.

## II. MATERIALS AND METHODS

### A. Experimental Setup

Seven healthy subjects (3 females, age range: 21-31 years) with normal or correct-to-normal vision participated in this study. They have received more than 10 years of formal English education. This study was reviewed and approved by the Institutional Review Board at Korea University (KUIRB-2019-0269-01).

*Research supported by Institute for Information & Communications Technology Promotion (IITP) grant funded by the Korea government (No. 2017-0-00451; Development of BCI based Brain and Cognitive Computing Technology for Recognizing User`s Intention using Deep Learning.)

J. Kalafatovich, and M. Lee are with the Department of Brain and Cognitive Engineering, Korea University, 145, Anam-ro, Seongbuk-gu, Seoul 02841, Republic of Korea (e-mail: jenifer@korea.ac.kr (J. Kalafatovich), minjilee@korea.ac.kr (M. Lee)).

S.-W. Lee is with the Department of Artificial Intelligence, Korea University, 145, Anam-ro, Seongbuk-gu, Seoul 02841, Republic of Korea (corresponding author: sw.lee@korea.ac.kr).

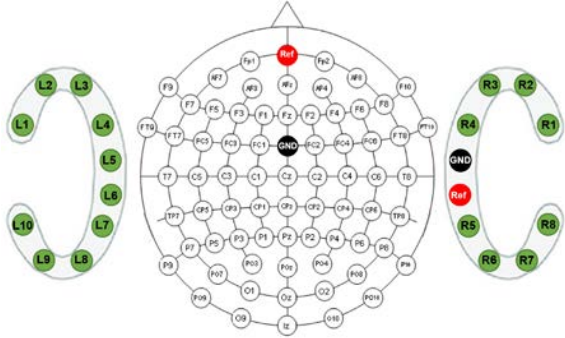

Figure 1. Location of electrodes of ear-EEG and scalp-EEG.

Scalp-EEG and ear-EEG data were simultaneously recorded. Scalp-EEG data was acquired using BrainAmp System at a sampling rate of 1,000 Hz; 62 electrodes were attached to the scalp according to the international 10–20 system. Reference and ground electrodes were placed at Fpz and FCz, respectively. Ear-EEG was recorded using Smarting System (mBrainTrain LLC, Serbia) and cEEGrid electrodes with 18 channels at a sampling rate of 500 Hz. Reference and ground electrodes were placed in the middle of the right ear cEEGrid. Two cEEGrid were attached to both ears of the subject, after which the EEG cap was put over the subject's head. Fig. 1 shows the locations of electrodes of ear-EEG and scalp-EEG. Prior signal measurement all channel impedances were set below 10 kΩ.

*B. Experimental Paradigm*

Subjects performed two phases of a memory task; (i) encoding; presentation of 250 English words in 5 lists of 50 word each, (ii) decoding; recognition test where subjects had to discern old and new words, in total 400 words that included 250 old words already presented in encoding phase and 150 new words firstly presented in the decoding phase were shown in random order. These phases were separated by an arithmetic distraction task, where subjects had to count backward during 20 m from 1,000 to zero in steps of seven. The distraction task aimed to avoid rehearsal of words [12].

Fig. 2 shows the experimental paradigm. For the encoding phase, a trial consisted of the presentation of fixation during 1 s, followed by a word during 2 s. Finally, subjects were asked to select the nature of the presented word (concrete or abstract) during 2 s. A list consisted of 50 trials, after which there was a black screen of 5 s. For the decoding phase, a trial consisted of the presentation of a fixation cross during 1 s, followed by a word (either new or old) during 2 s. After, subjects were asked to select a confidence scale from 1 (certain new) to 4 (certain old), in line with if the word was presented or not during the encoding phase. Trials were separated by a black screen of 1 s.

All words were randomly selected from a pool of 3,000 most commonly used English nouns according to Oxford University. The experimental paradigm was implemented using Psychophysics Toolbox in MATLAB software.

*C. Signal Processing*

EEG signals (ear and scalp-EEG) were down-sampled to 250 Hz and filtered between 0.5 and 40 Hz using a fifth-order Butterworth filter to reduce noise due to head and muscle movements. Prediction was done using signals of the encoding phase, for which data were separated in later successful remembered and forgotten trials depending on the selected confidence scale (if a word was presented in the encoding phase and subjects choose 1 or 2 it was considered forgotten). Each trial was epoch regarding stimulus onset into pre-stimulus and on-going stimulus (-1,000 and 1,000 ms) to compare time segment importance in memory process. All EEG channels were used for the analysis.

Feature extraction was performed using CSP. Previous studies had reported differences in frequency bands related to later remembered items in a memory task [16, 17], therefore FBCSP with theta [4-8 Hz], alpha [8-12 Hz], beta [12-30 Hz], and gamma bands [30-40 Hz] was also implemented. For classification LDA algorithm was applied.

Additionally, CNN was implemented. For the input of the CNN, raw EEG signals were bandpass filtered using same frequency bands as in FBCSP, resulting in a 4 dimensions tensor (trials × frequency band × time × channels). A kernel of 30 × N° electrodes was used to segment temporal signals and convolutional operation was applied with a stride of 25 (N° of electrodes varied for scalp- EEG and ear-EEG, 62 and 18 channels respectively). Rectifier linear unit (ReLU) was used as activation function. Additionally, dropout was applied to the input of the convolutional layer with a rate set at 0.25 to avoid overfitting. The convolutional layer generates 20 features maps that were the input to a fully connected layer with softmax activation. Cross entropy loss was used as loss function and all parameters were optimized

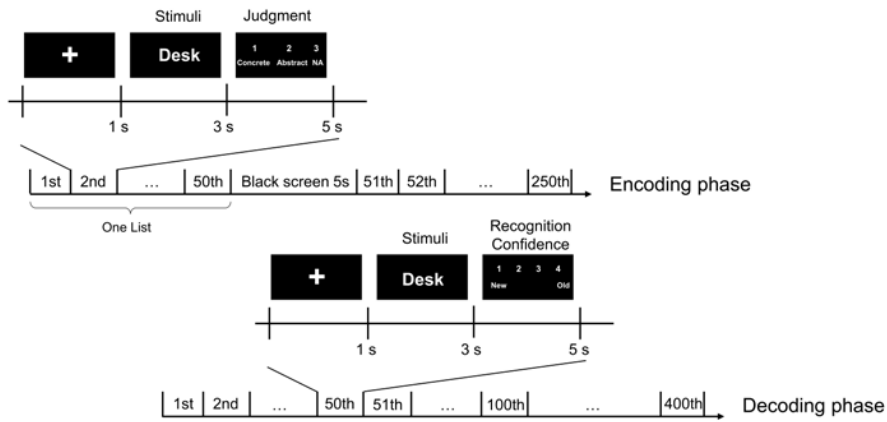

Figure 2. Experimental paradigm consisted in two phases: encoding (up) and decoding (down) phase, separated by a distraction task (20 m).

TABLE I. CNN ARCHITECTURE

| Layer | Operation | Kernel size | Feature maps |
|---|---|---|---|
| 1 | Dropout (p=0.25) Convolution Batch Normalization | - (30, N° electrodes) - | - 20 20 |
| Output | Fully Connected | - | 2 |

via Adam method. Table I shows the CNN architecture.

For evaluating the models, 10-fold-cross validation was used [19]. Data were divided into 10 groups; for each of the groups, the model was trained using the remaining 9 groups and evaluated taking the group as a test set. Test set accuracies were averaged and reported in the result section.

### D. Statistical Analysis

We performed the Kruskal-Wallis test to investigate the differences in the prediction performance of different classification methods across conditions (remembered and forgotten words during pre-stimulus and on-going stimulus). For post-hoc tests, Wilcoxon rank-sum tests were performed. All significance levels were set at $p = 0.05$.

## III. RESULTS

### A. Brain Activation during Memory Task

Fig. 3 shows power spectra difference between condition (remembered and forgotten trials) for pre and on-going stimulus across theta, alpha, beta, gamma bands. Significant changes over different brain regions were presented depending on the analyzed frequency band.

For theta band, there was a change over pre-frontal sites during pre-stimulus and over temporal sites during on-going stimulus. For alpha band, changes were localized over right temporal and parietal sites during pre-stimulus, and occipital sites during on-going stimulus. Analysis of beta band showed a difference over pre-frontal and occipital sites during pre-stimulus; and over occipital sites during on-going stimulus. Additionally, during pre-stimulus, there was a decrease in gamma band over pre-frontal and temporal sites, while an increase over pre-frontal sites during on-going stimulus. Difference between conditions was considerable, therefore, we use filtered raw EEG signals in four frequency bands (theta, alpha, beta, and gamma bands) to predict memory retrieval performance using CNN.

### B. Prediction of Memory Retrieval

Table II shows the prediction accuracies of different methods using ear-EEG and scalp-EEG data during the pre- and on-going stimuli. Using CNN, for scalp-EEG, average prediction accuracy achieved 74.59 ± 6.41% and 72.77 ± 9.04% using pre-stimulus and on-going stimulus segments, respectively. For ear-EEG, prediction accuracy achieved 74.06 ± 5.67% and 69.53 ± 9.04% using pre and on-going stimulus segments, respectively. There was no significant difference between pre and on-going stimuli. Additionally, no significant difference was found when comparing accuracies from ear-EEG and scalp-EEG for any method.

When applying CSP, accuracies whereas the following: For scalp-EEG, average prediction accuracy achieved 60.75 ± 8.28% and 59.09 ± 7.25% using pre-stimulus and on-going stimulus segments, respectively. For ear-EEG, average prediction accuracy achieved 60.73 ± 6.07% and 58.08 ± 5.75% using pre-stimulus and on-going stimulus segments, respectively. Statistical analysis showed no significant differences in predictive performance.

When applying FBCSP, accuracies whereas follow: For scalp-EEG, average accuracy achieved 63.41 ± 8.62% and 64.02 ± 8.51% using pre- and on-going stimuli, respectively. For ear-EEG, average prediction accuracy achieved 62.57 ± 7.32% and 57.66 ± 9.63% using pre-stimulus and on-going stimulus segments, respectively. There was no significant difference between pre and on-going stimuli.

### C. Comparison between Classifiers

The significant difference was found when comparing pre-stimulus accuracies across all methods for ear-EEG ($p = 0.039$) and scalp EEG ($p = 0.017$). Additionally, for scalp-EEG, significant difference was found when using on-going stimulus segments ($p = 0.021$). We also showed significant differences when compared to different methods accuracies with CNN. CNN accuracies were higher than CSP and LDA (pre-stimulus: $p = 0.009$ and 0.011, on-going stimulus: $p = 0.011$ and 0.020 for ear-EEG and scalp-EEG, respectively). CNN accuracies were higher than FBCSP and LDA, however significant difference was found only when comparing CNN performance and FBCSP during pre-stimulus segments ($p = 0.015$ and 0.026 for ear-EGG and scalp-EEG, respectively). CSP and FBCSP were applied to both datasets. FBCSP achieved higher prediction accuracy than CSP, except in on-going stimulus segments analysis for ear-EEG with no significant difference between them.

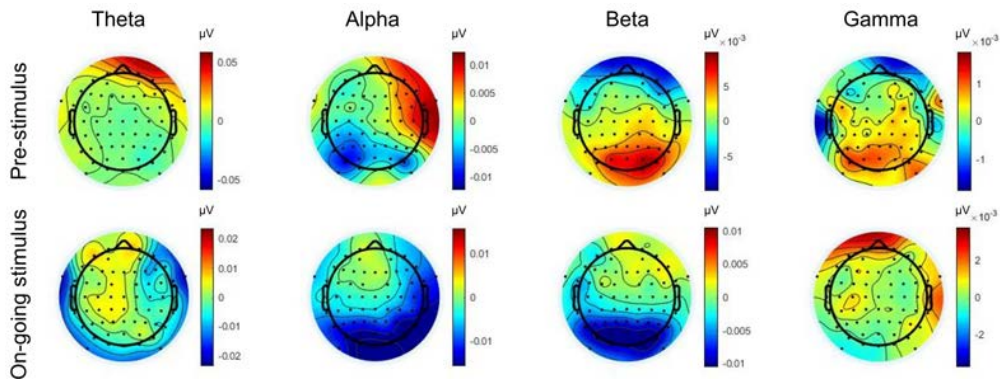

Figure 3. Power spectra difference between conditions of scalp-EEG during pre and on-going stimulus across different frequency bands.

TABLE II. PREDICTION ACCURACIES OF DIFFERENTE METHODS USING EAR-EEG AND SCALP-EEG

| Feature Extraction | Classifier | Pre-stimulus (%) | | On-stimulus (%) | |
|---|---|---|---|---|---|
| | | Ear-EEG | Scalp-EEG | Ear-EEG | Scalp-EEG |
| CNN[a] | | 74.06 ± 5.67 | 74.59 ± 6.41 | 69.53 ± 9.04 | 72.77 ± 9.04 |
| CSP[b] | LDA[d] | 60.73 ± 6.07 ($p = 0.009$) | 60.75 ± 8.28 ($p = 0.011$) | 58.08 ± 5.75 ($p = 0.011$) | 59.09 ± 7.25 ($p = 0.021$) |
| FBCSP[c] | | 62.57 ± 7.32 ($p = 0.015$) | 63.41 ± 8.62 ($p = 0.026$) | 57.66 ± 9.63 ($p = 0.091$) | 64.02 ± 8.51 ($p = 0.312$) |

a. Convolutional neural network, b. Common spatial pattern, c. Filter-bank CSP, d. Linear discriminant analysis.
*p*-values when comparing to all prediction accuracies to CNN.

## IV. DISCUSSION

Scalp EEG signals has been widely used in different studies [20, 21], including those related to memory [1-2]. However, we showed that it is possible to predict which items of a memory task are going to be remembered or not using ear-EEG with similar accuracy than scalp EEG (no significant difference between both). This can be justified by the relevance of temporal areas and spectral features in memory studies [13]. Highest prediction accuracy was obtained when applying the CNN, additionally, FBCSP performed better than CSP, except when analyzing on-going segment for ear-EEG. Previous studies reported significant changes in theta band over temporal sites associated with the formation of memory during pre-stimulus [14] and relevant increase activity of theta and low beta over frontal and temporal sites related to subsequent memory effects during on-going stimulus [12]. Therefore, the low prediction accuracy when using FBCSP for ear-EEG data can be due to the long distance from ear to task-related brain regions, our results showed the importance of different brain regions when analyzing frequency bands power spectral.

One limitation of this study is the number of trials used for the prediction analysis. Additionally, due to the physical characteristics of ear-EEG electrodes (cEEGrid electrodes), electrodes TP9 and TP10 of scalp-EEG were unable to measure, resulting in the measurement of 62 electrodes.

In conclusion, we aimed to predict the performance of a memory task using ear-EEG. Our findings showed a more practical way to predict performance of memory retrieval, due to similar accuracies obtained used ear-EEG and scalp-EEG, which can be very useful for the development of a brain-computer interface to enhance memory abilities. By predicting whether a stimulus is going to be more likely to be remembered or not, we can modulate the moment of stimulus presentation and increase memory performance [7].